\documentstyle[prl,aps,twoside]{revtex}

\begin{document}

\draft
\title{General criteria for the stability of uniaxially ordered states
of Incommensurate-Commensurate Systems}

\author{V. Danani\'{c}}  
 \address{ Department of Physics, Faculty of Chemical Engineering and 
 Technology,  University of Zagreb\\
Maruli\'{c}ev trg 19, 10000 Zagreb, Croatia}
\author{A. Bjeli\v{s} }
 \address{Department of Theoretical Physics, Faculty of Science, University 
 of Zagreb\\
POB 162, 10001 Zagreb, Croatia}
\maketitle

\begin{abstract}

Reconsidering the variational procedure for uniaxial systems modeled by
continuous free energy functionals, we derive new general conditions for
thermodynamic extrema. The utility of these conditions is briefly
illustrated on the models for the classes I and II of 
incommensurate-commensurate systems.

\end{abstract}

\pacs{05.70.Ce, 64.70.Rh, 82.60.-s}
%\narrowtext

\bigskip 

Numerous materials which are under intense investigations in the contemporary 
condensed matter physics are thermodynamically one-dimensional. The well-known
examples are various uniaxial materials with incommensurate and commensurate
orderings \cite{blc} and quasi one-dimensional conductors with charge or spin 
density wave instabilities \cite{gg}. Order parameters for such systems are
generally multicomponent, $\mbox{\bf u} = (u_1, u_2, ...., u_N)$,  and depend 
on a single spatial variable $x$. The principal task is then to find
thermodynamically stable configurations $\mbox{\bf u}_c(x)$, those which 
minimize the free energy functional ${\cal F}$. Since the latter is the 
one-dimensional integral, it is tempting to treat this variational problem 
as an equivalent to the standard classical mechanical one \cite{au}, with the
roles of time variable, vectors in the N-dimensional mechanical configuration
space, action functional and Lagrangian attributed to $x, \mbox{\bf u},
{\cal F}$ and $f$ respectively, the latter being the free energy density.  

In the present Letter we do not follow this widely accepted attitude, but start
from two obvious, yet substantial, differences between these two 
variational schemes. The first one is present in the very extremalization
procedure. In contrast to the classical mechanical trajectories, the realizable
solutions of the Euler-Lagrange (EL) equations for thermodynamic problems
follow after an additional extremalization with respect to the initial 
(or boundary) conditions.  The second difference concerns the content of the 
free energy densities. In the most interesting models for 
incommensurate-commensurate (IC) systems, including the basic ones, they
contain either terms linear in the first derivatives $\mbox { \bf u}^{'}=
(u_{1}', u_{2}', ...., u_{N}')$, or terms with higher derivatives 
$\mbox{\bf u}^{(j)}\equiv \frac{\partial^{j} {\bf u}}{\partial x^{j}}\, (j>1)$
(or both), in contrast to the standard mechanical Lagrangians 
which do not contain analogues of such terms. 

Starting from the first observation, we reformulate the procedure of
thermodynamic extremalization, and derive, under assumptions specified below,
the following necessary conditions for any thermodynamic extremum 
$\mbox{\bf u}_c$:
  
{\em Condition A;} 
\begin{equation}
\frac{1}{L}\int_{0}^{L}\left[\sum_{\alpha=1}^{N}\sum_{j=1}^{n}
\,j\,u_{c,\alpha}^{(j)}\frac{\partial f}{\partial u_{c,\alpha}^{(j)}}
-x \frac{\partial f}{\partial x} \right]\, dx=0,
\label{gc}
\end{equation}
where $n$ is the  order of highest derivative of $\mbox{\bf u}$ present in the
free energy functional, and $L$ is the length of the 
system taken in the thermodynamic limit $L \rightarrow \infty$.
In particular, for free energy densities which do not depend 
explicitly on $x$ the  condition  (\ref{gc}) reduces to the simple  equality
\begin{equation}
F_c  + H = 0, 
\label{T}
\end{equation}
where $F_c$ is the averaged value of free energy and $H$ is the integral 
constant which has the meaning of Hamiltonian in the equivalent classical 
mechanical problem (but does not have a direct physical meaning in the 
thermodynamic counterpart).

{\em Conditions B;}
\begin{equation}
\frac{1}{L}\int_{0}^{L}\sum_{j=0}^{n}
u_{c,\alpha}^{(j)}\frac{\partial f}{\partial u_{c,\alpha}^{(j)}}\, dx=0
\label{intgcu}
\end{equation}
where $\alpha = 1, ..., N$.

The ensuing discussion will show that in the case of thermodynamic functionals 
of the standard "mechanical" form the conditions (\ref{gc}) and (\ref{intgcu})
are of almost trivial meaning. They however have far-reaching implications
just in IC models, for which, as was already pointed out, free energy densities
depend in more complex ways on derivatives $\mbox{\bf u}^{(n)}$. These 
conditions also appear to be a powerful tool in the numerical determination of
phase diagrams, particularly for systems with nonintegrable free energy 
functionals.

In order to derive the {\em Conditions A} and {\em B}
we start from the general expression for the free energy functional 
\begin{equation}
{\cal F} = \frac{1}{L}\int_{0}^{L}
f\left[\mbox{\bf u}(x),\mbox{\bf u}'(x),\mbox{\bf u}''(x),...,
\mbox{\bf u}^{(n)}(x); x\right]dx,
\label{F}
\end{equation}
where $f$ is an analytical function of its arguments, bounded from below.
Each  thermodynamic extremum ${\bf u}_c(x)$ of this functional has to obey 
the variational condition $\delta{\cal F}(\{{\bf u}_c\}) =0$, equivalent to the 
Hamilton variational principle in classical mechanics. This necessary 
condition leads to the EL equations
\begin{equation}
 \sum_{j=0}^{n}(-1)^{j}\frac{d^{j}}{dx^{j}}\frac{\partial f}
 {\partial u_{\alpha}^{(j)}}=0 \,\,\,\,\, \,\,\ (\alpha=1,...,N),
\label{EL} 
\end{equation}
equivalent to the Lagrange equations in classical mechanics. The solutions of  
the EL equations (\ref{EL}) form a set $\{\mbox{\bf u}(x;\cal{A})\}$ which 
generally depends on $2 n N$ continuous parameters $(a_1, ...,  a_{2nN})\equiv
\cal{A}$. There is a freedom in the definition of the parameters ${\cal A}$,
the most usual choices being initial conditions [$\mbox{\bf u}(x_0), 
\mbox{\bf u}'(x_0),..., \mbox{\bf u}^{(2n -1)}(x_0)$] where $x_0$ is an 
arbitrary initial spatial position, and boundary conditions 
[$\mbox{\bf u}(x_1), \mbox{\bf u}'(x_1), ... 
\mbox{\bf u}^{(n-1)}(x_1); \mbox{\bf u}(x_2), \mbox{\bf u}'(x_2), ... 
\mbox{\bf u}^{(n-1)}(x_2)$] where $x_1$ and $x_2$ are arbitrary end points.
In classical mechanics these two choices correspond to the Newton and the 
Hamilton (variational) axiomatizations, respectively. Thermodynamic extrema,
including  thermodynamically stable configurations for which 
$\delta^2  {\cal F} \geq 0$, are those members of 
the set $\{\mbox{\bf u}(x;\cal{A})\}$ which extremalize the free energy 
${\cal F}(\{\mbox{\bf u};\cal{A})\})$ as a function of the parameters 
$\cal{A}$. This additional property completes, together with the EL equations 
(\ref{EL}), the sufficient condition for thermodynamic extrema. In particular,
a configuration which fulfills the conditions $\delta {\cal F} = 0$ and
$\delta^2  {\cal F} \geq 0$ is thermodynamically stable only if it is also a
minimum in the set $\{\mbox{\bf u}(x;\cal{A})\}$.

The dependence of ${\cal F}(\{\mbox{\bf u};\cal{A})\})$ on the parameters 
${\cal A}$ is generally intricate. It may be at least partly nonanalytic,
as is usually the case for the functionals (\ref{F}) with nonintegrable EL 
equations \cite{bb}, and in particular for those with free energy densities 
$f$ which are explicitly $x$-dependent. Thus, there is no efficient general
way to extract local extrema of ${\cal F}$ from the set 
$\{\mbox{\bf u}(x;\cal{A})\}$. However, we can now conveniently reformulate 
the above proposition that the thermodynamic extrema follow from the 
succession of the first order variation (\ref{EL}) and the extremalization 
with respect to the parameters ${\cal A}$, into an equivalent, and again 
sufficient, requirement that the solutions of the EL equations are 
thermodynamic extrema if they are local extrema in the set 
$\{\mbox{\bf u}(x)\}$ of {\em all} configurations allowed by the
functional (\ref{F}). By this enlargement of the set within which we are 
looking for the local extrema $\mbox{\bf u}_{c}(x)$, we get a freedom to 
choose arbitrarily (and suitably) the parameters with respect to which the 
set $\{\mbox{\bf u}(x)\}$ is analytic and corresponding extremalizations 
reduce to simple differentiations. This freedom will be here partly exploited, 
by making two choices of continuous parameters which will lead to the 
{\em Conditions A} and {\em B}.

The first continuous parameter is introduced in the  following way. 
Let us take one thermodynamic extremum, $\mbox{\bf u}_{c}(x)$,
and define a set of functions $\{\mbox {\bf u}(x; q)\}$ by 
\begin{equation}
\mbox{\bf u}(x; q) \equiv \mbox{\bf u}_{c}(qx).
\label{q}
\end{equation} 
The free energy functional (\ref{F}) for this set becomes
a function of $q$ given by
\begin{equation}
{\cal F}(\{\mbox{\bf u}(x; q)\}) \equiv F (q) =
\frac{1}{qL}\int_{0}^{qL}
f\left[\mbox{\bf u}_{c}(z),q\mbox{\bf u}_{c}'(z),q^{2}\mbox{\bf u}_{c}''(z),
....,q^{n}\mbox{\bf u}_{c}^{(n)}(z); q^{-1}z\right]\,dz,
\label{Fq}
\end{equation}
with $z\equiv qx$ and $\mbox{\bf u}_{c}^{(j)}(z) \equiv \partial 
\mbox{\bf u}_{c}^{j}(z)/\partial z^{j}$. The requirement that 
$\mbox{\bf u}_{c}(x)$ is an extremum in the set $\{\mbox {\bf u}(x; q)\}$ is 
expressed by
\begin{equation}
[\partial F(q)/\partial q]_{q=1}=0,
\label{q1}
\end{equation}
provided $F(q)$ is a smooth function of $q$ for $q\cong 1$. Let us also 
take the thermodynamic limit $L \rightarrow \infty$ and assume that 
$F(q)$ then does not depend on $L$ [up to the corrections of the order
$\cal{O}$$(1/L)$]. Under these assumptions, which will be critically 
examined later on, $F(q)$ may depend on $q$ only through the density 
$f$ in Eq.\ref{Fq}. The latter is an analytic function of $q$ since it is 
analytic with respect to $\mbox{\bf u}',... , \mbox{\bf u}^{(n)}$ by 
assumption. The derivative $\partial F(q) /\partial q$ is then well defined  
and the requirement (\ref{q1}), applied onto the function (\ref{Fq}),
gives the {\em Condition A}, Eq.\ref{gc}.

The further simplification takes place for the functionals (\ref{F}) in 
which the free energy density does not depend explicitly on $x$.  Then, like 
in classical mechanics, there exists an integral constant (Hamiltonian),
\begin{equation}
 H=-f+\sum_{\alpha=1}^{N}\left[\sum_{i=1}^{n}
 u_{\alpha}^{(i)}\frac{\partial f}{\partial u_{\alpha}^{(i)}}-
  \sum_{i=2}^{n}\sum_{j=0}^{i-2}
  (-1)^{j}u_{\alpha}^{(i-j-1)}\frac{d^{j+1}}{dx^{j+1}}
  \frac{\partial f}{\partial u_{\alpha}^{(i)}}\right], 
\label{H}
\end{equation}
for each solution of the EL equations (\ref{EL}). Using the obvious 
identity $H=\frac{1}{L}\int_{0}^{L}H\,dx$,  and the identity
\begin{equation}
\frac{1}{L}\int_{0}^{L}\left( u^{(k)}\frac{d^{l}}{dx^{l}}g\right)dx=
 \frac{1}{L}\left[\sum_{m=0}^{l-1}(-1)^{m}u^{(k+m)}
                                \frac{d^{l-m-1}}{dx^{l-m-1}}g\right]_{0}^{L} +
            (-1)^{l}\frac{1}{L}\int_{0}^{L}\left(u^{(k+l)}g\right)dx
\label{ident2}                         
\end{equation} 
which follows after $l$ successive partial integrations of the left-hand
side, one reduces the expression (\ref{H}) to 
\begin{equation}
H=-F_c
 -\frac{1}{L}\sum_{\alpha=1}^{N}\left[ 
\sum_{i=2}^{n}\sum_{j=0}^{i-2}
\sum_{k=0}^{j}(-1)^{j+k}u_{\alpha}^{(i-j+k-1)}\frac{d^{j-k}}{dx^{j-k}}
     \frac{\partial f}{\partial u_{\alpha}^{(i)}}\right]_{0}^{L} 
+ \frac{1}{L}\int_{0}^{L}\sum_{\alpha=1}^{N}
\sum_{j=1}^{n}\,j\,u_{\alpha}^{(j)}\frac{\partial f}
{\partial u_{\alpha}^{(j)}}\,dx.     
\label{HF}
\end{equation}
Here $g$ in Eq.\ref{ident2} is identified with 
$\partial f/\partial u_{\alpha}^{(i)}$ from Eq.\ref{H}, and 
$F_c \equiv {\cal F}(\{{\bf u}_c\})$. The second term on the right hand 
side in Eq.\ref{HF} is negligible in the limit $L \rightarrow \infty$, 
provided $\mbox{\bf u}_c (x)$ and its derivatives are finite. All 
thermodynamically stable extrema have this property since $f$ is bounded 
from below. The third term vanishes for each thermodynamic extremum due to 
the condition (\ref{gc}). The expression (\ref{HF}) thus reduces to the 
{\em Condition A}, Eq.\ref{T}. 

The equality (\ref{T}) is the consequence of the invariance of the functional 
(\ref{F}) with respect to translations in $x$, and of its non-invariance with
respect to the changes of $x$-scale. Like in classical mechanics, the former
invariance ensures the existence of the integral constant $H$ and the 
degeneracy of the solutions of EL equations with respect to the choice of 
"initial position" $x_0$.  The number of parameters on which the set 
$\{\mbox{\bf u}(x;\cal{A})\}$ explicitly depends is then $2 n N -1$. Note
that for all nontrivial functionals (\ref{F}) one has  $N \geq 1$ and 
$n \geq 1$, so that $2 n N -1 \geq 1$. In the simplest nontrivial case  
$N = n = 1$ the set ${\cal A}$ has one parameter, i. e.,  just $H$. 

For functionals (\ref{F}) with an explicit $x$-dependence of $f$, the
insertion of the EL equations (\ref{EL}) into the expression (\ref{gc}) 
leads to the relation $F_c = - H(L)$, where $H(L)$ is given by the, now 
$x$-dependent, expression (\ref{H}) at $x=L$. 
Since the right-hand side in this relation depends on L, it is inconsistent 
with at least one of two assumptions on the analyticity of $F(q)$ specified 
below Eq.\ref{q1}. We come to the conclusion that whenever the
free energy density depends explicitly on $x$, all thermodynamic extrema 
are isolated nonanalytical points of the corresponding functional (\ref{F})
with respect to changes of $x$-scale. This fundamental property is the
reason why the {\em Condition A} does not hold for such functionals.

Our second choice of continuous parameters from the set $\{\mbox{\bf u}(x)\}$ 
is defined by the scaling $u_{\alpha} \rightarrow s_{\alpha}u_{\alpha}$ for
any $1\leq \alpha \leq N$. The steps equivalent to those specified by 
Eqs.\ref{q}-\ref{q1} can be repeated now for each $\alpha$ for which 
$F(s_{\alpha})$ is a smooth function. The corresponding conditions
\begin{equation}
[\partial F(s_{\alpha})/\partial s_{\alpha}]_{s_{\alpha}=1} = 0
\label{salpha}
\end{equation}
then reduce to the {\em Conditions B}. Performing partial integrations and 
inserting EL equations (\ref{EL}) into  Eq.\ref{intgcu}, one finally gets 
the conditions
\begin{equation}
\frac{1}{L}\left[\sum_{l=0}^{n-1} u_{\alpha}^{(l)} 
\sum_{j=0}^{n-l-1} (-1)^{j} \frac{d^{j}}{d x^{j}}\,
\frac {\partial f}{\partial u_{\alpha}^{(j+1+l)}}\right]_{0}^{L}\,  = \, 0,
\label{gcu}
\end{equation}
which are constraints on the boundary values of the thermodynamic 
extrema. Note that the boundary ("surface") terms are here the leading ones, 
in contrast to the {\em Condition A} in which the analogous terms are only 
negligible ${\cal O}(1/L)$ corrections  to the finite volume terms of the 
order ${\cal O}(L^0)$. Obviously, any periodic solution of EL equations 
satisfies the conditions (\ref{gcu}). To this end it suffices to take into 
account corrections of the order ${\cal O}(1/L)$ coming from the boundary 
terms in {\em Conditions A and B}, in particular a correction which adjusts 
the period to be a divisor of $L$ with an integer ratio. No analogous 
adjustment for the quasiperiodic and nonperiodic solutions is apparent. The 
conditions (\ref{gcu}) are therefore expected to represent restrictive 
constraints on these solutions as possible candidates for thermodynamic 
configurations. 

The extremalization of the thermodynamic functional (\ref{F}) with respect
to the parameter set ${\cal A}$, and its non-invariance with respect to the
transformations $x \rightarrow qx $ and  $u_{\alpha} \rightarrow
s_{\alpha}u_{\alpha}$ in particular, become short of physical justification
when transposed to its mechanical counterpart.  For free energy densities 
which have the form of conservative Lagrangians one has $n = 1$, and the 
derivatives $u_{1}', u_{2}', ...., u_{N}'$ enter only through a positive 
definite quadratic form ("kinetic energy"). The criterion (\ref{T}) then 
singles out only equilibrium points (homogeneous configurations) $\mbox{\bf u}
= \mbox{\bf cte}$ as possible extrema. For such solutions the condition
(\ref{q1}) is trivially fulfilled, since $\mbox{\bf u}_{c}(qx)$ and the
corresponding free energy $F(q)$ does not depend on $q$. The same is 
true for the {\em Conditions B} which reduce to 
$[u_{\alpha}u^{'}_{\alpha}]_{0}^{L} = 0$.

As was announced in the introduction, the utility of the {\em Conditions 
A} and {\em B}\,  becomes apparent for the functionals (\ref{F}) which have 
richer dependences on the derivatives of ${\bf u}$ and allow for the 
$x$-dependent stable configurations.  For illustrations we take the basic 
models for the classes I and II of IC systems \cite{blc}, defined by \cite{mbh}
\begin{equation}
f = \frac{1}{2} (\phi' - \delta)^{2} - V(\phi)
\label{I}
\end{equation}
and \cite{hlsm}
\begin{equation}
f = (u'')^{2} -  (u')^{2} + \lambda u^2 + \frac{1}{2} u^4
\label{II}
\end{equation}
respectively. 

The decisive term in the model (\ref{I}) is the Lifshitz invariant 
$\delta \phi'$. $\phi$ is the phase variable, so that $V(\phi + 2\pi) = 
V(\phi)$, the simplest choice being the sine-Gordon model with a single 
Umklapp term, $V(\phi) \propto \cos p\phi$, where $p$ is an integer. The 
problem (\ref{F}, \ref{I}) is entirely solvable \cite{mbh,tol}, since the 
corresponding EL equation is integrable and the set ${\cal A}$ has one 
parameter, e. g., $H$. Here we show how the {\em Condition A} enables an 
elegant derivation and an original interpretation of the solution. The {\em 
Condition A} for the functional (\ref{F}, \ref{I}) reduces to 
\begin{equation}
2\pi\delta = I_c \equiv \int_{0}^{2\pi}\phi'_{c}(\phi) d\phi 
= \int_{0}^{2\pi}\sqrt{2[-F_c + \delta^{2}/2 - V(\phi)]} d\phi.
\label{sl}
\end{equation}
The determination of  the thermodynamic phase diagram, i. e., of the 
dependence of $F_c$ on the control parameters present in the model (\ref{I}),
is thus reduced to the calculation of the integral $I_c$. The relation
(\ref{sl}) also states that for a thermodynamic extremum the 
corresponding mechanical action variable is just equal to the Lifshitz 
parameter $\delta$! The dependence of the period $P$ of the stable 
configuration on control parameters follows from the known relation for
mechanical systems with one degree of freedom, $P = \partial I_c/\partial H$
\cite{arn}. Finally, the corresponding configuration $\phi_{c}(x)$ follows 
from the quadrature of the EL equation with an {\em already determined value 
 of $H$}. Thus, using the equality (\ref{T}) we avoid a more tedious procedure 
used in the analyses of the models (\ref{I}) \cite{mbh,tol}, namely, the entire 
integration of the EL equation (with free $H$) followed by the minimization 
of the free energy $F(H)$ as a function of $H$.

Since the transformation (\ref{q}) already exhausts the freedom in the choice 
of variational parameters ${\cal A}$ for the model  (\ref{I}), the {\em 
Condition B} which is now given by $[\phi (\phi' - \delta)]_{0}^{L} = 0$, 
cannot be an additional constraint,  but may only reproduce some already 
derived property of the extremum $\phi_c$. This condition states 
that the configuration  $\phi_c$ has a slope $\phi' = \delta$ at the points $x = 
0,  P,  2P, ..., NP $,  where $P$ is a period and $N$ is a large (macroscopic) 
integer. It indeed follows independently from the EL equation and the 
condition (\ref{sl}).

Various criteria suggest \cite{db1} that the model (\ref{II})
is nonintegrable due to the presence of the second derivative of the 
real order parameter u. Very probably
$H = (u'')^2 - (u')^2 - 2 u' u''' - \lambda u^2 - \frac{1}{2} u^4$
is the only integral constant among three parameters in the set ${\cal A}$. 
The {\em Condition A} now reads
\begin{equation}
 \int_{0}^{L}\left[ 2 (u_{c}^{''})^2 - (u_{c}^{'})^2\right] dx = 0.
\label{IIT}
\end{equation}
Without using this condition, we have minimized numerically the functional
(\ref{F}, \ref{II}) in the Fourier basis and showed that the phase diagram 
contains an enumerable set of metastable periodic solutions with homogeneous 
domains connected by sinusoidal segments \cite{db1}. The subsequent check 
\cite{db2} verifies that all these solutions satisfy the condition 
(\ref{IIT}). Furthermore, by using it, one significantly facilitates the
numerical calculation of (meta)stable configurations for the model
(\ref{II}). Namely, the search for local minima in the Fourier basis
gives, as a rule, continuous families of periodic configurations. In order to
find the proper thermodynamic configurations within one family it suffices to 
determine zeros of the diagonal quadratic form of Fourier components to which
the left-hand side of equation (\ref{IIT}) reduces. By this we directly
confirm that the obtained configuration  satisfies the EL equation and 
determine its period.  The more detailed presentation of this procedure for
the model (\ref{II}) and its various extensions is given elsewhere \cite{db2}. 

Applying the {\em Condition  B} to the model (\ref{II}) we obtain an 
additional constraint on the boundary points, 
\begin{equation}
 [u' u'' - u(u' + u''')]_{0}^{L} = 0,
\label{IIu}
\end{equation}
which, together with the arguments given below Eq.\ref{gcu},
reinforces the expectation based on the
independent numerical analysis \cite{db1} that all thermodynamic extrema
of the problem (\ref{II}) are very probably periodic.
Note that by conditions (\ref{IIT}) and (\ref{IIu}) we have fixed 
two out of three parameters from the set ${\cal A}$ for the problem (\ref{II}).
Very probably ${\cal F}$ is not analytic for any choice of the remaining
third variational parameter, in close connection with the nonintegrability of 
the EL equation and the corresponding chaotic structure of the portrait in 
the phase space $(u, u', u'', u''')$.  

Having these and other \cite{db2} examples in mind, we connect the 
limitations of the present method with the degree of the nonintegrability 
of a given functional by the following conjecture: larger is the number of 
missing integral constants (in the classical mechanical sense), smaller is 
the number of analytic conditions for the thermodynamic extrema (like those 
given by {\em Conditions A} and {\em B}).

In conclusion, necessary conditions for uniaxial thermodynamic extrema are
obtained from the extremalization with respect to space and order parameter
scales. This procedure proves to be feasible for the free energy densities 
which are not 
explicitly dependent on the space variable. In particular, we show that in 
this case the sum of the averaged free energy and the integral constant 
(Hamiltonian) vanishes for each thermodynamic extremum. Besides their 
general significance, the present results will be certainly of practical use in  
analytical and numerical analyses of particular models for uniaxial systems. 

We acknowledge discussions with M. Latkovi\'{c}. The work is supported by 
the Ministry of Science and Technology of the Republic of Croatia through the 
project no. 119201.

\end{document}